\documentclass[aps,prl,twocolumn,superscriptaddress,groupedaddress,amsmath,amssymb]{revtex4-1}
\usepackage{subfigure}
\usepackage{graphicx}
\usepackage{dcolumn}
\usepackage{bm}
\usepackage{xcolor}
\usepackage{float}
\usepackage{booktabs}
\definecolor{aa}{RGB}{0,0,139}
\newcommand{\BESIIIorcid}[1]{\href{https://orcid.org/#1}{\hspace*{0.1em}\raisebox{-0.45ex}{\includegraphics[width=1em]{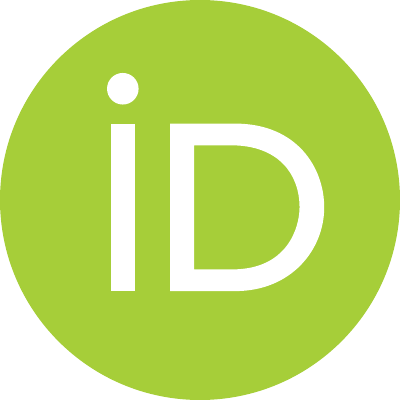}}}}

\usepackage[colorlinks,urlcolor=aa,linkcolor=blue,anchorcolor=aa,citecolor=aa]{hyperref}

\usepackage[mathlines]{lineno}
\usepackage{multirow}
\usepackage{enumerate}
\usepackage{amsmath}
\usepackage{relsize}
\usepackage{makecell}
\def\babar{\mbox{\slshape B\kern-0.1em{\smaller A}\kern-0.1em
    B\kern-0.1em{\smaller A\kern-0.2em R}}}

\makeatletter
\renewcommand\footnote[1]{%
    \stepcounter{footnote}%
    \protected@xdef\@thefnmark{\thefootnote}%
    \@footnotemark
    \@footnotetext{#1}%
}
\makeatother

\def\MeanBFa{\ensuremath{(0.98 \pm 0.16_{\rm{stat.}} \pm 0.22_{\rm{syst.}})\%}}
\def\BFnoetap{\ensuremath{(2.97 \pm 0.23_{\rm stat.} \pm 0.14_{\rm sys.})\%}}

\def\BFall{\ensuremath{(3.79 \pm 0.22_{\rm stat.} \pm 0.14_{\rm sys.})\%}}
\def\BFetap{\ensuremath{(0.82 \pm 0.08_{\rm stat.} \pm 0.02_{\rm sys.})\%}}

\def\BFetapp{\ensuremath{(3.66 \pm 0.36_{\rm stat.} \pm 0.09_{\rm sys.} \pm 0.08_{\rm PDG})\%}}

\begin{document}
\preprint{APS/123-QED}



\title{{\bf \boldmath Observation of  $D_s^+ \to a_0(980)^+f_0(500)$ in the Amplitude Analysis of $D_s^+ \to \pi^+ \pi^0 \pi^0 \eta$}}

\author{
\begin{small}
\begin{center}
M.~Ablikim$^{1}$\BESIIIorcid{0000-0002-3935-619X},
M.~N.~Achasov$^{4,c}$\BESIIIorcid{0000-0002-9400-8622},
P.~Adlarson$^{81}$\BESIIIorcid{0000-0001-6280-3851},
X.~C.~Ai$^{87}$\BESIIIorcid{0000-0003-3856-2415},
C.~S.~Akondi$^{31A,31B}$\BESIIIorcid{0000-0001-6303-5217},
R.~Aliberti$^{39}$\BESIIIorcid{0000-0003-3500-4012},
A.~Amoroso$^{80A,80C}$\BESIIIorcid{0000-0002-3095-8610},
Q.~An$^{77,64,\dagger}$,
Y.~H.~An$^{87}$\BESIIIorcid{0009-0008-3419-0849},
Y.~Bai$^{62}$\BESIIIorcid{0000-0001-6593-5665}
% ...
\\
\vspace{0.2cm}
(BESIII Collaboration)\\
\vspace{0.2cm} {\it
$^{1}$ Institute of High Energy Physics, Beijing 100049, People's Republic of China\\
% ...
$^{4}$ Budker Institute of Nuclear Physics SB RAS (BINP), Novosibirsk 630090, Russia\\
% ...
$^{31}$ INFN Sezione di Ferrara, (A)INFN Sezione di Ferrara, I-44122, Ferrara, Italy; (B)University of Ferrara, I-44122, Ferrara, Italy\\
% ...
$^{39}$ Johannes Gutenberg University of Mainz, Johann-Joachim-Becher-Weg 45, D-55099 Mainz, Germany\\
% ...
$^{62}$ Southeast University, Nanjing 211100, People's Republic of China\\
% ...
$^{64}$ State Key Laboratory of Particle Detection and Electronics, Beijing 100049, Hefei 230026, People's Republic of China\\
% ...
$^{77}$ University of Science and Technology of China, Hefei 230026, People's Republic of China\\
% ...
$^{80}$ University of Turin and INFN, (A)University of Turin, I-10125, Turin, Italy; (B)University of Eastern Piedmont, I-15121, Alessandria, Italy; (C)INFN, I-10125, Turin, Italy\\
$^{81}$ Uppsala University, Box 516, SE-75120 Uppsala, Sweden\\
% ...
$^{87}$ Zhengzhou University, Zhengzhou 450001, People's Republic of China\\

\vspace{0.2cm}
$^{\dagger}$ Deceased\\
% ...
$^{c}$ Also at the Novosibirsk State University, Novosibirsk, 630090, Russia\\
% ...

}

\end{center}
\vspace{0.4cm}
\end{small}
}

\vspace{0.4cm}
\begin{abstract}
  We report the first observation of the decay $D_s^+ \to \pi^+\pi^0\pi^0\eta$ in a
  data set corresponding to an integrated luminosity of 7.33~fb$^{-1}$,
  collected in $e^+e^-$ collisions by the BESIII detector at center-of-mass
  energies between 4.128 and 4.226~GeV. An unexpectedly large branching fraction $\mathcal{B}( D_s^+ \to a_0(980)^+ f_0(500), a_0(980)^+ \to \pi^+\eta, f_0(500)\to \pi^0\pi^0$) = \MeanBFa \ is measured with a significance exceeding \( 10\sigma \), offering new constraints on the internal structure of light scalar mesons. 
  The dominant intermediate process is $D_s^+ \to a_1(1260)^+\eta, a_1(1260)^+\to \rho(770)^+\pi^0$ with a branching fraction of $(1.77 \pm 0.21_{\rm stat.} \pm 0.12_{\rm syst.})\%$.     The isospin symmetry has been validated to  the decays of $a_1(1260)^+\to \rho(770)^0\pi^+$ and $a_1(1260)^+\to \rho(770)^+\pi^0$. 
  Moreover, the measured $\mathcal{B}(D_s^+\to \pi^+\pi^0\pi^0\eta|_{\rm{non}-\eta^\prime})$ = \BFnoetap \ reduces the undetected $D_s^+ \to \eta X$ decay branching fractions to (0.1 $\pm$ 3.1)\%. 

\end{abstract}

\pacs{Valid PACS appear here}
\maketitle



Charmed hadronic decays provide a unique laboratory for studying weak and strong interactions in the non-perturbative regime of Quantum Chromodynamics~(QCD), which is inaccessible to perturbative methods. The copious decay products of charmed mesons also provide an ideal platform to investigate the nature of light mesons. The low-lying axial-vector meson $a_1(1260)$ and the
scalar mesons below 1~GeV/$c^2$---$a_0(980), f_0(980)$, and $f_0(500)$---remain poorly understood. Their internal structures and even their basic properties continue to show significant experimental uncertainties~\cite{ParticleDataGroup:2024cfk},
and all their decay channels are listed as
``seen" without quantitative information on their branching fractions~(BFs). 
The $a_1(1260)$ and $a_0(980)/f_0(980)$ masses are close to the strange meson pair thresholds: $K\bar{K}^*$  for the former, and $K\bar{K}$ the latter. Their widths suggest that both states may arise from a common underlying mechanism.
Theoretical studies~\cite{Hsiao:2019ait, Yu:2021euw, Hsiao:2023qtk, Belle:2002gzj, Roca:2005nm, Geng:2008ag, Aceti:2016yeb, Dai:2020vfc, Clymton:2022jmv} have also shown that the $\bar{K}K^*$ and $\bar{K}K$ channels may
play essential roles in producing the $a_1(1260)$ and $a_0(980)/f_0(980)$, respectively,
leading to their frequent interpretations as compact tetraquark or molecular states.

\begin{figure}[tp!]
\begin{center}
   \flushleft
   \begin{minipage}[t]{8.8cm}
      \includegraphics[width=0.492\linewidth]{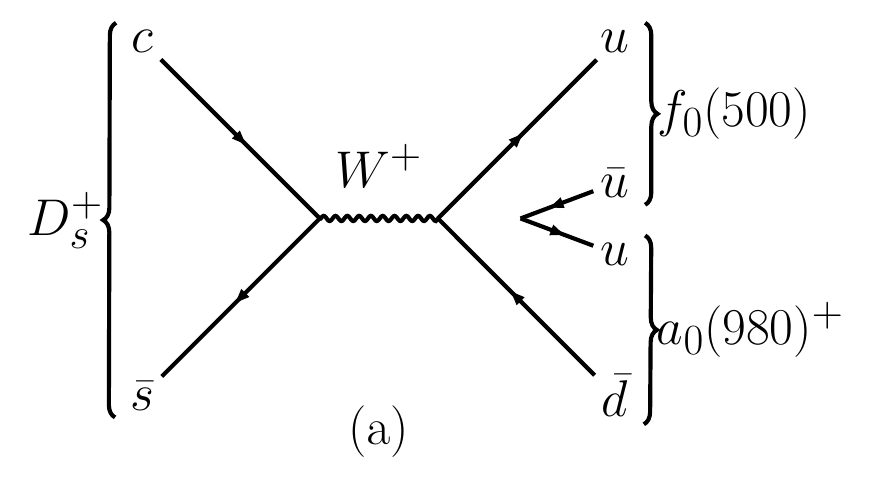}
      \includegraphics[width=0.492\linewidth]{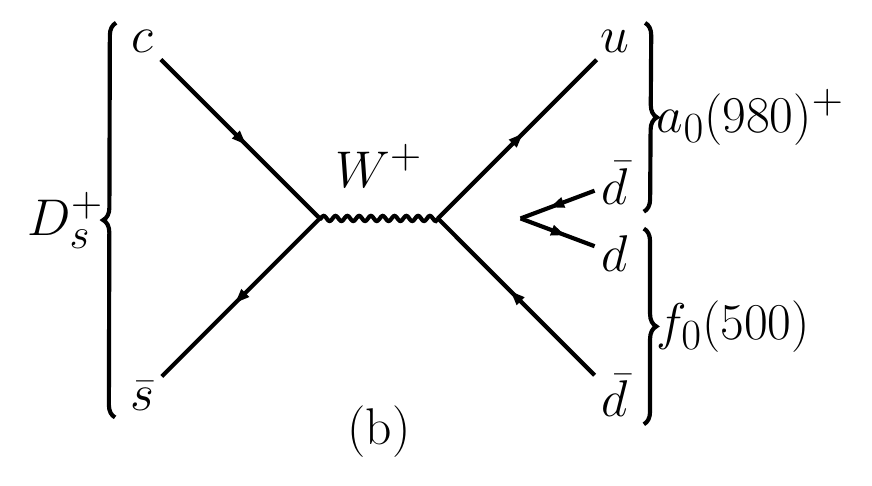}\\
       \includegraphics[width=0.492\linewidth]{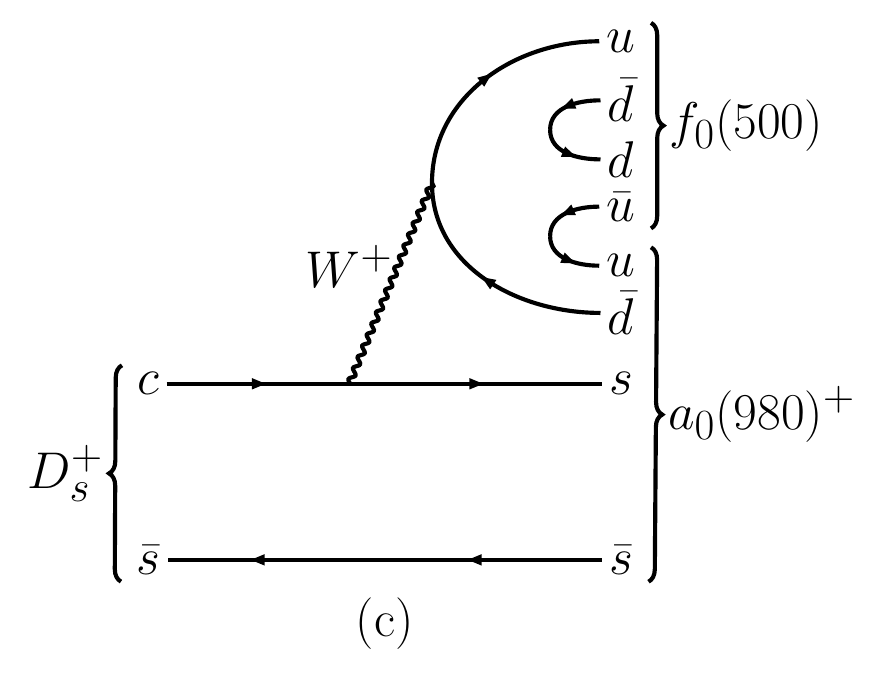}
     \includegraphics[width=0.492\linewidth]{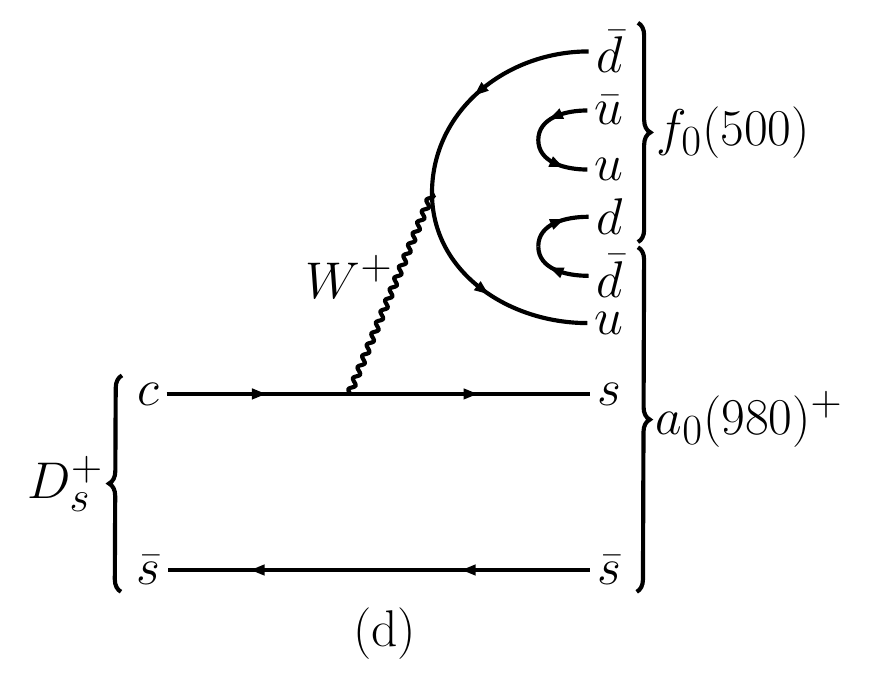}\\ 
   \end{minipage}
   \caption{Topological diagrams of the decay $D_s^+ \to a_0(980)^+ f_0(500)$, where the $a_0(980)^+$ and $f_0(500)$ are represented in (a-b) as a two-quark state and in (c-d) as a tetraquark state.}
\label{topologya0}
\end{center}
\end{figure}

Understanding charmed meson decay dynamics has been significantly
improved through systematic amplitude analyses, which enable the extraction of
quasi-two-body decay parameters from multi-body final
states~\cite{BESIII:2023htx}. 
Numerous $D_{(s)} \to SP$ and $D_{(s)} \to SV$ decays, where $S$, $P$ and $V$ 
denote scalar, pseudoscalar and vector mesons, have been measured~\cite{ParticleDataGroup:2024cfk} and studied theoretically~\cite{Cheng:2002mk, Cheng:2002ai, Boito:2008zk, Cheng:2022vbw, Dai:2023jix} to investigate the exotic nature of scalar particles below 1~GeV/$c^2$. So far, no $D_{s}\to SS$ decays have been observed. In a tetraquark scenario, the hadronization of each $S$ requires an additional quark-antiquark pair, which makes
$D_{s}\to SS$ more sensitive to the production mechanisms of scalar particles, such as final state interaction.
The intermediate process $D_s^+ \to a_0(980)^+ f_0(500)$ of $D_s^+ \to \pi^+\pi^0\pi^0\eta$ is expected to proceed through pure weak annihilation~(WA), as illustrated in Figs.~\ref{topologya0}(a), (b). The BF should be suppressed by the small $a_0(980)^+$ decay constant under the assumption that the scalar mesons are conventional two-quark mesons. 
However, if scalar mesons are tetraquarks, this decay mode may receive a contribution from additional external $W$-emission amplitudes, shown in Figs.~\ref{topologya0}(c), (d), which could significantly
enhance its BF~\cite{Cheng:2024zul}. Determining the BF of
$D_s^+ \to a_0(980)^+ f_0(500)$ therefore offers a valuable opportunity
to probe the internal structure of the $a_0(980)$ and $f_0(500)$. It is
important to note that compared to $D_s^+ \to \pi^+\pi^+\pi^-\eta$, there is no
interference from the $\rho^0(770)$ in $D_s^+ \to \pi^+\pi^0\pi^0\eta$, making it more
suitable for the study of $D_s^+ \to SS$ decays.

Experimental measurements of the processes $D_{(s)}^+ \to a_1(1260) P$, remain limited and exhibit discrepancies with theoretical calculations~\cite{Cheng:2003bn}. In the previous study of $D_s^+\to \pi^+\pi^+\pi^-\eta$~\cite{BESIII:2021aza},  the intermediate process $D_s^+ \to a_1(1260)^+\eta, a_1(1260)^+ \to \rho(770)^0\pi^+$ has been observed with a 
 BF of $(1.73\pm0.14_{\rm stat.}\pm0.08_{\rm syst.})\%$. Similarly, the unobserved decay mode $D_s^+ \to \pi^+\pi^0\pi^0\eta$ is expected to proceed dominantly through
$D_s^+ \to a_1(1260)^+\eta, a_1(1260)^+ \to \rho(770)^+\pi^0$. The information
of these decays will be crucial for validating isospin symmetry in
$a_1(1260)^+$ decays and add further constraints for a more comprehensive understanding of the properties of $a_1(1260)$. 


Previous BESIII analyses have observed the decays $D_s^+\to a_0(980)^{0}\pi^{+}, a_0(980)^{0}\to \pi^{0}\eta$ and $D_s^+\to a_0(980)^{+}\pi^{0}, a_0(980)^{+}\to \pi^{+}\eta$ with mutually consistent BFs of $(1.46 \pm 0.15_{\rm stat}\pm0.23_{\rm sys})$\%~\cite{BESIII:2019jjr}.
Ref.~\cite{Yu:2021euw} predicts the same for
$D_s^+\to a_0(980)^{+}\rho^{0}(a_0(980)^{0}\rho^{+})$. While the $D_s^+\to a_0(980)^{+}\rho^{0}$ has been observed in $D_s^+ \to \pi^+\pi^+\pi^-\eta$~\cite{BESIII:2021aza}, this
work reports the first search for the corresponding
$D_s^+ \to a_0(980)^0\rho^+$ decay channel.

In this Letter, we perform the first amplitude analysis and BF measurement of
$D_s^+ \to \pi^+\pi^0\pi^0\eta$ using the data sets collected with the BESIII
detector with a total integrated luminosity of
7.33~fb$^{-1}$~\cite{BESIII:2022dxl}. We report the first observation of the
$D_s \to SS$ decay $D_s^+ \to a_0(980)^+f_0(500)$. Charge-conjugate states and
exchange symmetry of two identical $\pi^0$ candidates are implied throughout this Letter.


A description of the design and performance of the BESIII detector can be found in Ref.~\cite{BESIII:2009fln}. 
Monte Carlo (MC) events are simulated with a {\sc{geant4}}-based~\cite{GEANT4:2002zbu} detector simulation software, which includes the geometric description~\cite{Huang:2022wuo} and the detector response. 
The inclusive MC samples generated with KKMC include open charm processes, initial state radiation~\cite{Kuraev:1985hb} production of vector charmonium(-like) states and the continuum processes incorporated in {\sc{kkmc}}~\cite{Jadach:1999vf, Jadach:2000ir}.  The open charm processes are generated using {\sc{conexc}}~\cite{Ping:2013jka}. Final state radiation is considered using {\sc{photos}}~\cite{Barberio:1993qi}.
All particle decays are modelled with {\sc
evtgen}~\cite{Lange:2001uf,Ping:2008zz} using BFs 
either taken from the
PDG~\cite{ParticleDataGroup:2022pth}, when available,
or otherwise estimated with {\sc lundcharm}~\cite{Chen:2000tv,Yang:2014vra}.

In the data samples, the $D_s^{\pm}$ mesons are produced mainly via $e^+e^- \to D_s^{*\pm}D_s^\mp \to \gamma D_s^\pm D_s^\mp$. Therefore, the double-tag (DT) method~\cite{MARK-III:1985hbd} is employed to perform the analysis, in which a single-tag (ST) candidate is reconstructed using nine hadronic decays: $D_{s}^{-}\to K_{S}^{0}K^{-}$, $D_{s}^{-}\to K^{+}K^{-}\pi^{-}$, $D^{-}_{s}\to K_S^{0}K^{-}\pi^{0}$, $D^{-}_{s}\to K^{+}K^{-}\pi^{-}\pi^{0}$, $D^{-}_{s}\to \pi^{+}\pi^{-}\pi^{-}$, $D^{-}_{s}\to \pi^{-}\eta_{\gamma\gamma}$, $D^{-}_{s}\to \pi^{-}\pi^0\eta_{\gamma\gamma}$, $D^{-}_{s}\to \pi^{-}\eta^\prime_{\pi^+\pi^-\eta_{\gamma\gamma}}$, and $D^{-}_{s}\to K^-\pi^-\pi^+$. The DT candidates are formed by selecting a $D_{s}^{+} \to \pi^+\pi^0\pi^0\eta$ decay in the side of the event recoiling against the $D_s^-$ meson. The selection criteria for the final state particles, the transition photon and the $D_s^-$ candidates are the same as in Ref.~\cite{BESIII:2023mie}.
To reduce combinatorial background, a nine-constraint (9C) kinematic fit is performed, incorporating the constraints of four-momentum conservation in the $e^+e^-$ center-of-mass system, the invariant mass constraints of the $\pi^0$, the $\eta$, and the tagged $D_s^-$ meson to their respective known masses~\cite{ParticleDataGroup:2022pth}, as well as constraints applied to either the $D_s^+$ or $D_s^-$ candidate along with the selected transition photon to the $D_s^{*+}$ mass~\cite{ParticleDataGroup:2022pth}. In cases where multiple candidates exist in an event, the one with the minimum $\chi^2$ value of the 9C kinematic fit is selected. A further kinematic fit including a constraint on the mass of the signal $D_s^+$ is performed to ensure that all events are within the phase space boundary, and the updated four-momenta by the kinematic fit are used for the amplitude analysis.

The events stemming from the decay $D_s^+ \to \pi^+\eta^\prime, \eta^\prime\to\pi^0\pi^0\eta$ are eliminated by excluding the events with an invariant mass of $\pi^0\pi^0\eta$ below $1.10~{\rm GeV}/c^2$ in the amplitude analysis. Events in which photons from $\pi^0$, $\eta$, and $D_s^{*+}$ are miscombined into a fake $\pi^0$ are filtered out by vetoing the invariant mass combinations falling within the range [0.10, 0.14]~${\rm GeV}/c^2$.
Additionally, the recoiling mass against $D_s^{*+}$, $M_{\rm rec}$, must be consistent with $D_s$, falling within the range [1.93, 1.98] GeV/$c^2$ and defined as
\begin{equation}
\small
    M_{\rm {rec}} = \sqrt{\left(E_{\rm cm}-\sqrt{|\vec{p}_{D_{s}^{+}\gamma}|^2c^2 + m^2_{D_s^{*}}c^4}\right)^2/c^4-|\vec{p}_{D_s^{+}}|^2/c^2},
\end{equation}
where $E_{\rm cm}$ is the center-of-mass energy, $\vec{p}_{D_{s}^{+}\gamma}$ is the sum of the momentum of signal $D_s^+$ and the transition photon, $\vec{p}_{D_{s}^{+}}$ is the momentum of signal $D_s^+$, and $m_{D_s^{*}}$ is the known $D_s^{*+}$ mass. 
We also require that the transition photon energy in the laboratory frame must be below 0.18 ${\rm GeV}$ according to the MC simulation. In total, 821 $D_s^+ \to \pi^+\pi^0\pi^0\eta$ event candidates are selected with a purity of $(58.3 \pm 1.7)\%$, as ascertained by fitting the invariant mass distribution of the signal $D_s^+$ candidates. The remaining backgrounds are mainly from non-$D_s^{*\pm}D_s^{\mp}$ open charm processes.

An unbinned maximum likelihood method is adopted in the amplitude analysis. The fit probability density function is the sum of the signal amplitude and a background function, with their coefficients corresponding to their fractions. The signal amplitudes and background functions are described in the following. 
The signal amplitude is parameterized with the isobar formulation in the covariant tensor formalism~\cite{Zou:2002ar}. 
The total signal amplitude $\mathcal{M}$ is expressed as a coherent sum of intermediate processes, given by $\mathcal{M} = \sum \rho_n e^{i\phi_n}\mathcal{A}_n$,  where $\rho_n e^{i\phi_n}$ represents the coefficient of the $n$-th amplitude with $\rho_n$ denoting its magnitude and $\phi_n$ its phase. The $n$-th amplitude $\mathcal{A}_n$ is defined as $\mathcal{A}_n = F_n^{r_1} F_n^{r_2}F_n^{D_s}S_nP_n^{r_1}P_n^{r_2}$. Here, $F_n^{D_s}$ and $F_n^{r_i}$ correspond to the  Blatt-Weisskopf barrier factors~\cite{Blatt_Weisskopf_1973} of the $D_s^+$ meson and the $i$-th intermediate state, respectively. The term $S_n$ represents the spin factor~\cite{Zou:2002ar},  while the $P_n^{r_i}$ denotes the propagator for the $i$-th resonance. In the multibody decays, $r_1$ and $r_2$ can form resonant or non-resonant systems and the amplitudes are constructed considering both possibilities.

The relativistic Breit-Wigner (RBW) function~\cite{Jackson:1964zd} is used to describe the propagator of the $a_1(1260)$ resonance, whose width is considered as a quasi-three-body decay and determined by integrating the squared amplitude over phase space~\cite{BESIII:2023qgj}.
The $\rho(770)$ resonance is parametrized by the Gounaris-Sakurai~(GS) line shape~\cite{Gounaris:1968mw}.
A coupled Flatt\'e formula is utilized for the $a_0(980)$ resonance, with parameters fixed to the values given in Ref.~\cite{BESIII:2016tqo}.
The $f_0(500)$  resonance is parametrized using a formula that was derived fitting $\pi \pi$ elastic scattering data~\cite{Bugg:1996ki}.
The $K$-matrix formalism in the $P$-vector approximation~\cite{Adler:1964um, Aitchison:1972ay, Anisovich:2002ij} is used to model the $\pi^0\pi^0~S$-wave, with contributions from $\pi\pi,K\bar{K}, \pi\pi\pi\pi, \eta\eta$, and $\eta\eta^\prime$ scattering processes. Due to the limited statistics available in this analysis, the $K$-matrix parameters are fixed to the available experimental results~\cite{Anisovich:2002ij, BaBar:2008inr, BaBar:2018cka}.

The masses and widths of the remaining intermediate resonances are taken from Ref.~\cite{ParticleDataGroup:2022pth}. The background shape is estimated with inclusive MC samples using the XGBoost package~\cite{Rogozhnikov:2016bdp,Liu:2019huh}. 

\begin{figure*}
  \begin{center}
    \includegraphics[width=0.7\textwidth]{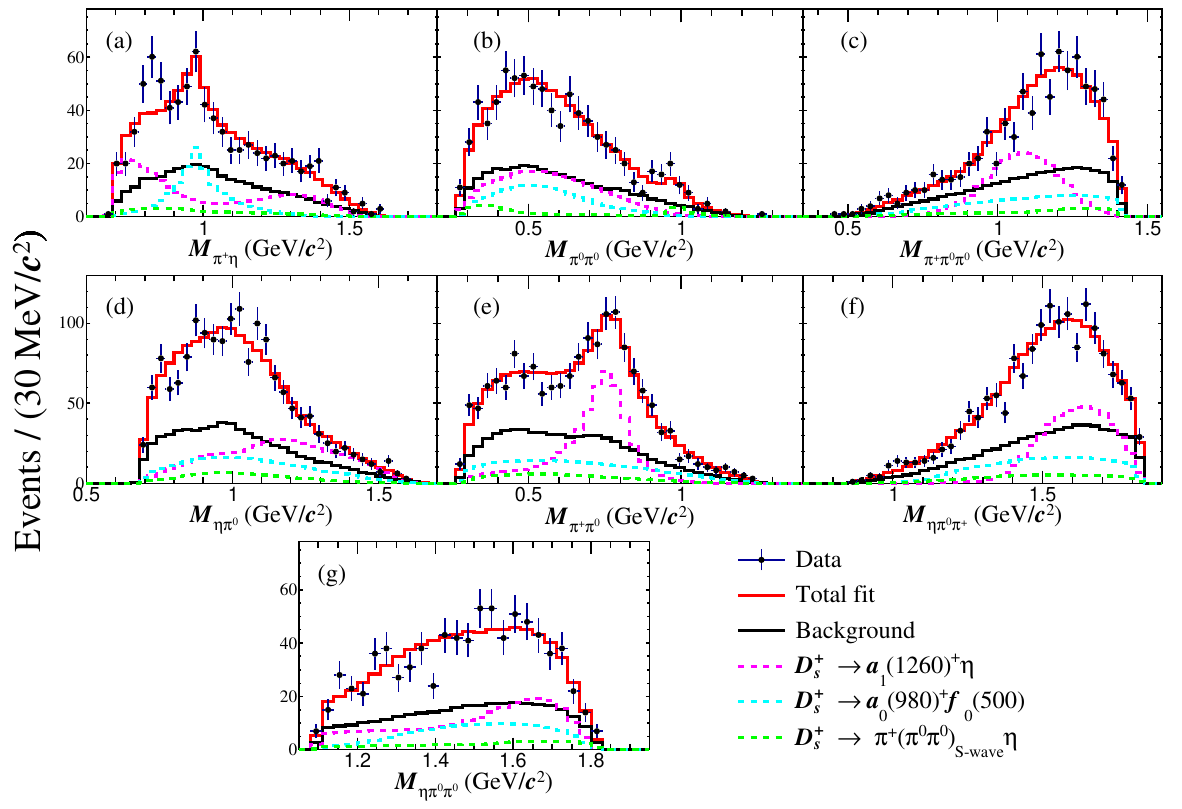}
    \caption{The projections of the fit are shown on (a) $M_{\pi^+\eta}$, (b) $M_{\pi^0\pi^0}$, (c) $M_{\pi^+\pi^0\pi^0}$, (d) $M_{\eta\pi^0}$, (e) $M_{\pi^+\pi^0}$, (f) $M_{\eta\pi^0\pi^+}$ and (g) $M_{\eta\pi^0\pi^0}$. The combinations of the two identical $\pi^0$s are added in (d), and (e), because of the exchange symmetry. 
    }
    \label{pwa:fitresult}
  \end{center}
\end{figure*}

In the fit process, the decay $D_s^+\to a_1(1260)^+\eta$, $a_1(1260)^+\to \rho(770)^+\pi^0$ is assumed for the initial amplitude model, with the magnitude and phase fixed to one and to zero, respectively, as a reference to the magnitude and phase of other states, owing to their evident presence in the invariant mass projections.
Then, all the possible combinations with different intermediate processes are tested and the model including the processes with statistical significance larger than $3\sigma$ is kept, where the statistical significance of each amplitude is calculated based on the change of the log-likelihood with and without this amplitude by taking the change of the degrees of freedom into account. 
The fitted invariant mass projections using the final model are shown in Fig.~\ref{pwa:fitresult} and the corresponding fitted phases, fitted fractions (FFs), and their statistical significances are listed in Table~\ref{pwa:fitresulttable}.
The FF of the $n$-th amplitude is calculated by 
\begin{equation}
{\rm FF}_n = \int{|\rho_n e^{i\phi_n}\mathcal{A}_n|^2}d\Phi_4/\int{|\mathcal{M}|^2}d\Phi_4,
\end{equation}
where $d\Phi_4$ is the volume element of the four-body phase space. The interference between amplitudes are given in the Supplemental Material. 
An enhancement in the data has been observed around 0.8 GeV$/c^2$ in the \( M(\pi^+\eta) \) spectrum. We try to fit this peak with additional amplitudes, however, given the limited statistics, the improvement is not significant (around 3$\sigma$).

\begin{table*}[tp!]
	\caption{Phases, FFs, BFs and statistical significances for the amplitudes. The first and second uncertainties are statistical and systematic, respectively. }
	\centering
	\setlength{\tabcolsep}{4.5pt}
	\renewcommand{\arraystretch}{1.2}
	\begin{tabular}{lc r@{ $\pm$ }c@{ $\pm$ }c c c}
	\hline
	\hline
		Amplitude  &Phase (rad) &\multicolumn{3}{c}{FF (\%)} &BF (\%) &Significance\\
		\hline
		 $D_s^{+}\to a_1(1260)^+\eta, a_1(1260)^+\to \rho(770)^+\pi^0$  &0.0 (fixed) &59.7 &5.2  &2.9 &1.77 $\pm$ 0.21 $\pm$ 0.12 &$>10\sigma$\\
		 \hline
		 $D_s^{+}\to a_0(980)^+f_0(500)$  &2.8 $\pm$ 0.2 $\pm$ {0.3} &33.1 &4.9  &6.8 &0.98 $\pm$ 0.16 $\pm$ 0.22&$>10\sigma$\\
		 \hline
		 $D_s^{+}\to \pi^+(\pi^0\pi^0)_{\mathcal{S}-\rm{wave}}\eta$ &-0.8 $\pm$ 0.2 $\pm$ {0.3} &9.4 &1.9  & 4.1 &0.28 $\pm$ 0.06 $\pm$ 0.13&4.5$\sigma$\\
    \hline
	\hline
    \end{tabular}
		\label{pwa:fitresulttable}
\end{table*}

The systematic uncertainties of the quantities fitted with the amplitude analysis are assigned as the difference between alternative fits and the nominal fit as follows. 
The systematic uncertainty associated with the resonance modeling is assessed by propagating the statistical uncertainties in the masses and widths~\cite{ParticleDataGroup:2022pth}, converting the lineshape of the $\rho(770)^+$ to the RBW form~\cite{Jackson:1964zd}, and replacing the lineshape of the $f_0(500)$ to the RBW function~\cite{Jackson:1964zd} with the fixed values of 526 MeV/$c^2$ for the mass and 534 MeV/$c^2$ for the width~\cite{BES:2004mws}.
A uniform distribution with the range of 4 $({\rm GeV}/c)^{-1}$ is assumed for the barrier radius. The systematic uncertainty is estimated by repeating the fit procedure while varying the radius by its 1$\sigma$, $\pm4/\sqrt{12} \approx \pm1\ ({\rm GeV}/c)^{-1}$.
The uncertainty associated with the detector acceptance difference between data and MC simulation is determined by reweighting the MC events in likelihood function according to the detector acceptance difference studied using samples of $e^+e^- \to K^+K^-\pi^+\pi^-(\pi^0)$ processes, as the same method in Ref.~\cite{BESIII:2020ctr}. 
The uncertainty caused by purity is studied by varying the purity within its statistical uncertainty, and the background shape is studied by varying the proportion of MC background components by $\pm30\%$ according to the statistical uncertainties of the cross sections of non-$D_s^{*\pm}D_s^{\mp}$ open charm mesons in the inclusive MC samples. 
The intermediate resonances with statistical significances less than $3\sigma$ are included in the fit one by one and the largest difference with respect to the baseline fit is taken as the systematic uncertainty. 
To estimate the potential biases in the fit process, 600 data-sized simulation samples are generated with the amplitude models from the nominal analysis and refitted by an identical procedure. All the fitted pull values that deviate from zero are assigned as the corresponding systematic uncertainties. The total uncertainties are determined by adding all the contributions in quadrature. The detailed results can be found in the Supplemental Material~\cite{Supplemental}.

The same tag modes as in the amplitude analysis are employed to perform the BF measurement. To account for the contribution from the decay $D_s^+ \to \pi^+ \eta^\prime, \eta^\prime\to \pi^0\pi^0\eta$, the cut on $M_{\pi^0 \pi^0 \eta}$ is removed. This contribution is identified by selecting events where $M_{\pi^0 \pi^0 \eta}$ lies within the $\eta^\prime$ mass range of [0.946, 0.970] GeV/$c^2$.
If there are multiple tag $D_s^-$ candidates for each tag mode, then the one with the recoiling mass against $D_s^-$, $M_{\rm rec0}$, closest to the known $D_s^{*+}$ mass is retained. The DT candidate with an average mass (($M_{\rm sig} + M_{\rm tag}$)/2 closest to the known $D_s^+$ mass~\cite{ParticleDataGroup:2022pth} is retained for each tag mode. 

\begin{figure}[htp]
  \begin{center}
    \includegraphics[width=0.23\textwidth]{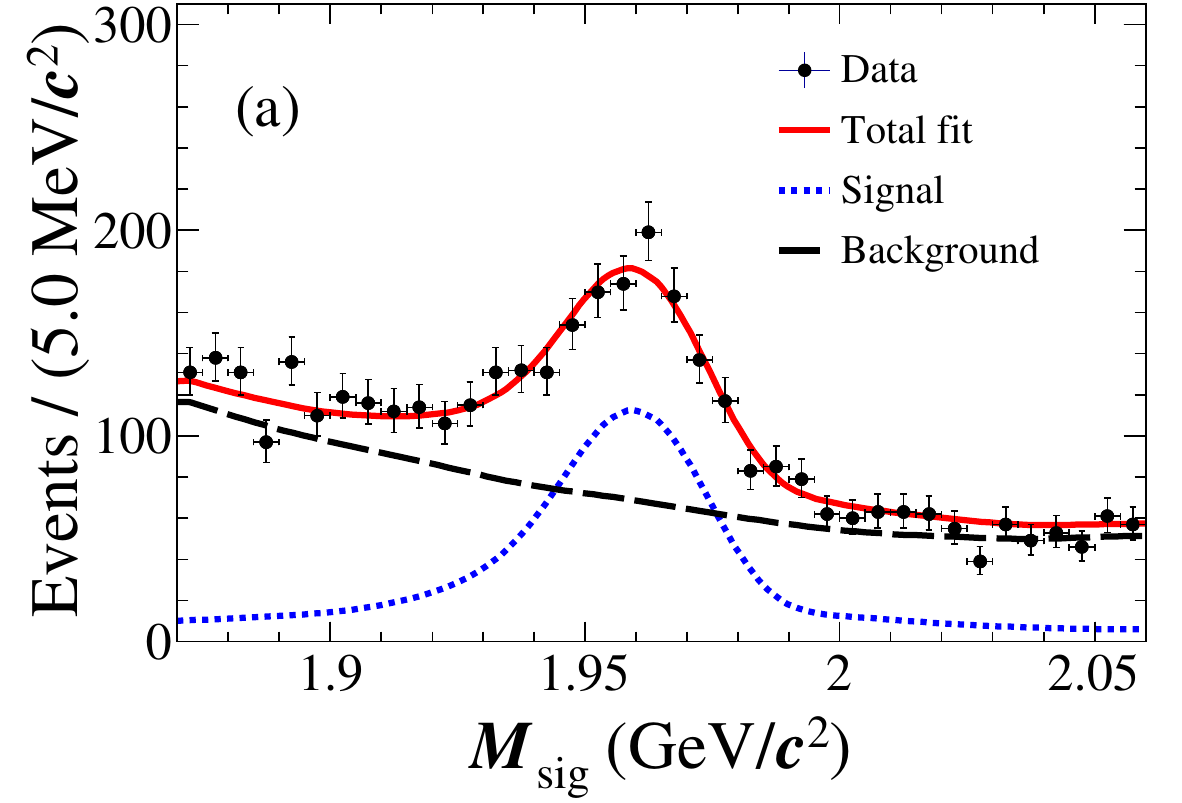}
    \includegraphics[width=0.23\textwidth]{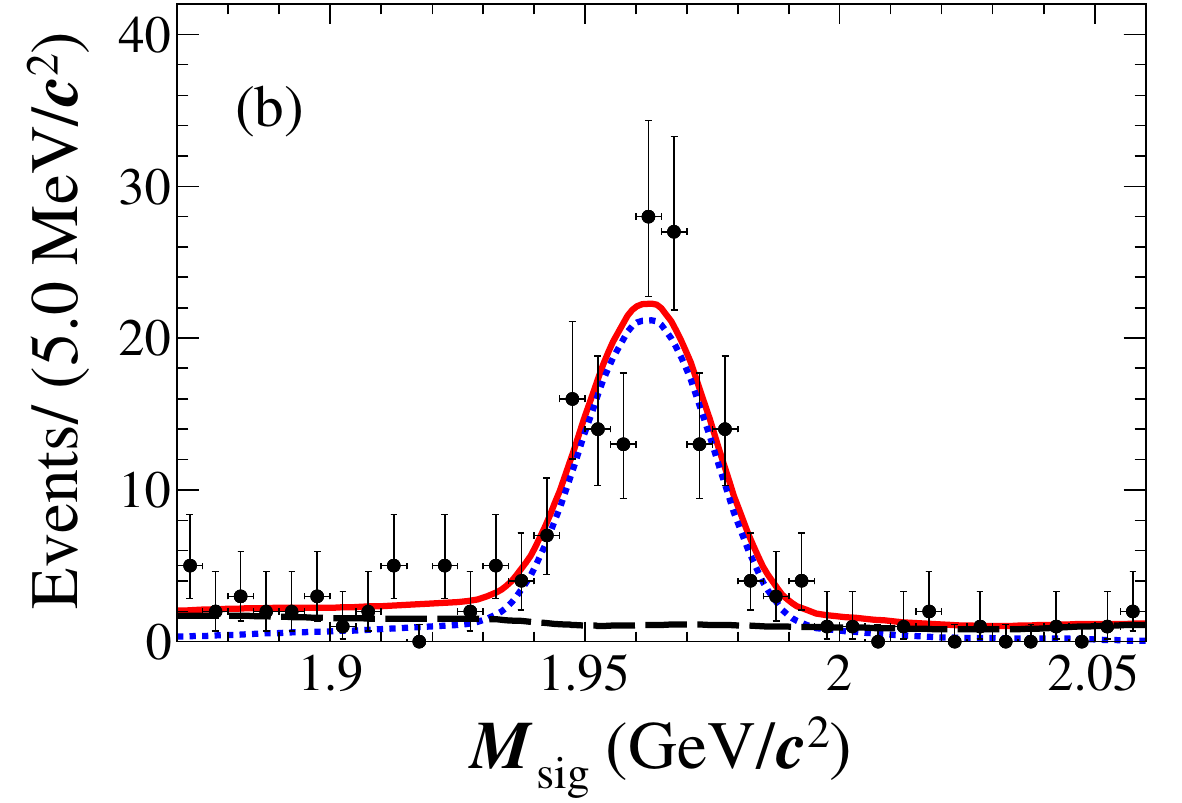}    
    \caption{Fits to the $M_{\rm sig}$ distributions of DT candidates for (a) all events and (b) $\pi^+\eta^\prime$. The data samples are represented by points with error bars, the total fit by the red line, and the fitted signal and background by the blue dotted and black dashed lines, respectively. 
    }
    \label{BF:fitresult}
  \end{center}
\end{figure}

A fit to the invariant mass of $D_s^\pm$ is performed to obtain the ST yields ($Y_{\rm tag})$ and DT yields ($Y_{\rm sig}$), together with the ST efficiencies ($\epsilon_{\rm tag}$) and DT efficiencies ($\epsilon_{\rm tag,sig}$) estimated with the corresponding signal MC samples, resulting in the relation $\mathcal{B}(D_s^+ \to \pi^+\pi^0\pi^0\eta) = (Y_{\rm sig}/\sum_i Y_{\rm tag}^i \epsilon_{\rm tag,sig}^i/\epsilon_{\rm tag}^i)$, where the index $i$ denotes the $i$-th tag mode. The ST fit results are the same as in Ref.~\cite{BESIII:2023mie}. The result of the fit to the invariant mass of the signal $D_s^+$ is shown in Fig.~\ref{BF:fitresult}. In the fit, the signal shape is the convolution of the simulated signal shape and a Gaussian function, while the background shape is described with the shape obtained from the inclusive MC samples. 
With fitted DT yields from the non-$\eta^\prime$ part being $Y_{\rm sig, all} = 1176\pm70$ and for the $\pi^+\eta^\prime$ part being $Y_{\rm sig, \pi^+\eta^\prime} = 143\pm15$, the BFs are determined to be $\mathcal{B}(D_{s}^{+} \to\pi^{+}\pi^{0}\pi^{0}\eta)$ = \BFall~and $\mathcal{B}(D_{s}^{+} \to\pi^{+}\eta^\prime, \eta^\prime \to \pi^{0}\pi^{0}\eta)$ = \BFetap~by dividing by the BFs of $\pi^0\to\gamma\gamma$ and $\eta \to \gamma \gamma$~\cite{ParticleDataGroup:2022pth}. The BF of the non-$\eta^\prime$ decay mode is obtained by subtracting the $\pi^+\eta^\prime$ contribution, which is determined to be $\mathcal{B}(D_{s}^{+} \to\pi^{+}\pi^{0}\pi^{0}\eta|_{\rm non-\eta^\prime})$ = \BFnoetap.

For the BF measurement, the relative uncertainty in the total number of ST $D_s^-$ mesons is assigned to be 0.5\%, accounting for the statistical fluctuation. The uncertainty related to the background shape in the fit to signal $D_s^+$ distribution is assigned by repeating the fit with the proportion of MC background components varied by $\pm30\%$. 
The systematic uncertainties related to $\pi^\pm$ tracking and particle identification as well as the reconstruction efficiencies of $\pi^0$ are estimated by studying $e^+e^- \to K^+K^-\pi^+\pi^-(\pi^0)$ events. 
The systematic uncertainty due to the amplitude analysis model is studied by varying the parameters in the amplitude analysis fit according to the covariance matrix. Adding all sources of uncertainties in quadrature gives a total systematic uncertainty of 3.7\% for $\mathcal{B}(D_s^+ \to \pi^+\pi^0\pi^0\eta|_{\rm non-\eta^\prime})$ and 2.8\% for $\mathcal{B}(D_s^+ \to \pi^+\eta^\prime,\eta^\prime\to\pi^0\pi^0\eta)$. The detailed results can be found in the Supplemental Material.

In summary, the decay $D_s^+ \to \pi^+\pi^0\pi^0\eta$ has been observed for the
first time, with BFs of $\mathcal{B}(D_{s}^{+} \to\pi^{+}\pi^{0}\pi^{0}\eta|_{\rm non-\eta^\prime})$ = \BFnoetap~and~$\mathcal{B}(D_{s}^{+} \to\pi^{+}\eta^\prime, \eta^\prime \to \pi^{0}\pi^{0}\eta)$ = \BFetap.
Utilizing $\mathcal{B}(\eta^\prime \to \pi^0\pi^0\eta)=(22.4 \pm 0.5)\%$ quoted from the PDG~\cite{ParticleDataGroup:2024cfk}, the BF $\mathcal{B}(D_s^+\to \pi^+\eta^\prime)$ is determined to be \BFetapp. 
This result reduces the inferred BF of unobserved $D_s^+ \to \eta X$ decays to (0.1 $\pm$ 3.1)\%.

Based on the results of the first amplitude analysis and the measured
$\mathcal{B}(D_{s}^{+} \to\pi^{+}\pi^{0}\pi^{0}\eta|_{\rm non-\eta^\prime})$,
the absolute BFs of intermediate states are calculated as
$\mathcal{B}_i = {\rm FF}_i \times \mathcal{B}(D_s^+ \to \pi^+\pi^0\pi^0\eta)$.
The decay of $D_s^+ \to a_0(980)^+f_0(500)$ is observed for the first time, with its BF determined to be \MeanBFa~with a significance exceeding 10$\sigma$. The obtained BF is unexpectedly large (of order $10^{-2}$), which may indicate a substantial tetraquark component in the $a_0(980)$ or significant contributions from final state interactions of triangle rescattering~\cite{Wang:2025mdn}. 
The BF of $D_s^+\to a_1(1260)^+\eta, a_1(1260)^+\to\rho(770)^+\pi^0$ is
determined to be $(1.77\pm0.21_{\rm stat}\pm0.12_{\rm syst})\%$, taking into
account the interferences between other amplitudes. Comparing with
$\mathcal{B}(D_s^+\to a_1(1260)^+\eta, a_1(1260)^+\to\rho(770)^0\pi^+) = (1.73\pm0.14_{\rm stat}\pm0.08_{\rm syst})\%$ measured by BESIII~\cite{BESIII:2021aza}, the isospin symmetry is
validated for the decays of $a_1(1260)^+\to \rho(770)^0\pi^+$ and $a_1(1260)^+\to \rho(770)^+\pi^0$. 
As for the $D_s^+ \to a_0(980)^0\rho(770)^+$ decay, the change in the log-likelihood corresponds to a significance of only 2.4$\sigma$, therefore its contribution with the current statistics is not significant.
Furthermore, we observe a structure around 0.8 GeV/$c^2$ in the $M_{\pi^+\eta}$
spectrum as shown in Fig.~\ref{pwa:fitresult}(a), which cannot be described by known resonances. Similar structures are also visible in the $D_s^+ \to \pi^+\pi^+\pi^-\eta$ decay~\cite{BESIII:2021aza}. These observations provide new input for theoretical models of scalar mesons, particularly regarding their possible tetraquark or molecular nature. Future studies with increased statistics~\cite{Achasov:2023gey} will provide deeper insights into these phenomena.
  
\textbf{Acknowledgement}

The BESIII Collaboration thanks the staff of BEPCII (https://cstr.cn/31109.02.BEPC) and the IHEP computing center for their strong support. This work is supported in part by National Key R\&D Program of China under Contracts Nos. 2023YFA1606000, 2023YFA1606704; National Natural Science Foundation of China (NSFC) under Contracts Nos. 12375070, 11635010, 11935015, 11935016, 11935018, 12025502, 12035009, 12035013, 12061131003, 12192260, 12192261, 12192262, 12192263, 12192264, 12192265, 12221005, 12225509, 12235017, 12342502, 12361141819; the Chinese Academy of Sciences (CAS) Large-Scale Scientific Facility Program; the Strategic Priority Research Program of Chinese Academy of Sciences under Contract No. XDA0480600; CAS under Contract No. YSBR-101; Joint Large-Scale Scientific Facility Funds of the NSFC and CAS under Contracts Nos. U2032108; Shanghai Leading Talent Program of Eastern Talent Plan under Contract No. JLH5913002; 100 Talents Program of CAS; The Institute of Nuclear and Particle Physics (INPAC) and Shanghai Key Laboratory for Particle Physics and Cosmology; ERC under Contract No. 758462; German Research Foundation DFG under Contract No. FOR5327; Istituto Nazionale di Fisica Nucleare, Italy; Knut and Alice Wallenberg Foundation under Contracts Nos. 2021.0174, 2021.0299, 2023.0315; Ministry of Development of Turkey under Contract No. DPT2006K-120470; National Research Foundation of Korea under Contract No. NRF-2022R1A2C1092335; National Science and Technology fund of Mongolia; Polish National Science Centre under Contract No. 2024/53/B/ST2/00975; STFC (United Kingdom); Swedish Research Council under Contract No. 2019.04595; U. S. Department of Energy under Contract No. DE-FG02-05ER41374


\bibliography{draft_CH_CL.bib}

\end{document}